\documentclass[10pt]{article}
\usepackage{amssymb}
\usepackage{amsmath}

\newcommand{\egy}[1]{\begin{equation}#1\end{equation}}

\begin{document}

\title{Edwards-Wilkinson surface over a spherical substrate: $1/f$ noise in the height fluctuations} 

\author{V.~Eisler and Sz.~Farkas \\ \\ Institute for Theoretical Physics,
E\"otv\"os University,\\ 1117 Budapest, P\'azm\'any s\'et\'any 1/a, Hungary}

\date{\today}

\maketitle

%\address{Institute for Theoretical Physics,
%E\"otv\"os University, 1117 Budapest, P\'azm\'any s\'et\'any 1/a, Hungary}

\begin{abstract}
We study the steady state fluctuations of an Edwards-Wilkinson type surface
with the substrate taken to be a sphere. We show that the height fluctuations
on circles at a given latitude has the effective action of a perfect Gaussian
$1/f$ noise, just as in the case of fixed radius circles on an infinite planar
substrate. The effective surface tension, which is the overall coefficient
of the action, does not depend on the latitude angle of the circles.

\end{abstract}

%\pacs{}

\section{Introduction}
It was long ago that the voltage fluctuations of a current-carrying resistor was
observed to have a power spectrum nearly proportional to the inverse of the
frequency \cite{Weissman}. Since then it has been realized that this so called $1/f$ noise
can be detected in a large variety of phenomena. Examples range from the velocity fluctuations of
moving interfaces \cite{movint}, the number of stocks traded daily \cite{stock}
to the spectra of speech and music \cite{music}.
Because of this apparent universality, the suggestion emerged
that there might be a generic underlying mechanism.
The problem has been extensively studied from a large number of viewpoints, but a
general description has not emerged so far.
\par
Antal \emph{et al.} have recently found \cite{Antal} a new aspect of the $1/f$
noise. Namely, they proved that the mean square fluctuations
(also called roughness) of periodic signals with Gaussian $1/f$ power
spectra are distributed according to one of the extreme value probability distributions,
the Fisher-Tippet-Gumbel (FTG) distribution \cite{FT}\cite{Gumbel}.
This means that if we can find physical
systems where Gaussian $1/f$ noise is present and periodic boundary
conditions are realized, then FTG statistics naturally emerges
in a measurement, thus again suggesting a possible reason for the generality.
\par
The periodic boundary conditions are of course strange restrictions
when we look for a physical realization.
One example was given in \cite{Antal} for a system which meets
the above requirements. If we consider a $d=2$ Edwards-Wilkinson (EW) surface on
an infinite planar substrate, and draw a circle on it with a given radius $R$,
then the probability of a height configuration is given by the effective action of
the desired Gaussian $1/f$ noise.
\par
Here we consider the generalization of the above example to spherical
substrates. A motivation for investigating this problem comes from the
increasing amount of accessible data on the cosmic microwave background (CMB)
radiaton's anisotropy.
Recently CMB measurements of the Wilkinson Microwave  Anisotropy Probe first year sky
survey gave a detailed sky map on the temperature fluctuations of the
radiation \cite{Bennett}.
These fluctuations are believed to be Gaussian, and their angular power spectrum
is an essential detail in the cosmological models.
The statistics of the present data are still not perfect, and one could improve
the reliability of conclusions by investigating the fluctuations on
objects of various shapes.
Here we investigate the case of circles on the sphere and we assume that the
EW fluctuations provide a simple model of the temperature anisotropy.
If we could prove that the fluctuations on circles
of this spherical substrate have the same properties as in the planar case, then a
comparison between the probability distribution of the measured data and
the FTG distriubtion would tell us about the appropriateness of our assumption.
We could also look at the decay of two point correlations of the data set,
which is in fact a more direct way of checking the EW model assumption.
However, we believe that the use of the distribution function might
give an effective alternative method, since here we have not only
an exponent, but a universal scaling function to compare.
\par
In our paper we present the derivation of the probability density functional
of a height configuration over the equator of the sphere.
We conclude that the effective action is equal to that of the
Gaussian $1/f$ noise, exactly as in the case of the planar substrate in
\cite{Antal}. Then we generalize our result to arbitrary circles with
latitude angle $\vartheta_0$.

\section{The model}

In EW model the probability $P[h]\sim e^{-S[h]}$ of a height configuration 
$h(\vartheta,\varphi)$
defined on the 2-sphere $S_2$ can be given through the free-field effective action
\egy{S\left[ h \right]=\frac{\sigma}{2}\int\limits_{S_2}
\left|\nabla h(\vartheta,\varphi)\right|^2\,\mathrm{d}\Omega\label{action},}
where $\sigma$ is the effective surface tension. In order to get the
probability that $h$ takes the value of some fixed $h_0$
on the equator we have to integrate over field values
at other points:
\egy{P\left[ h_0 \right]=Z^{-1}\int_
{h\left(\frac{\pi}{2},\varphi\right)=h_0\left(\varphi\right)}
\mathcal{D}h\,e^{-S\left[ h \right]}\label{funcint1},}
where $Z$ is a normalization coefficient, and equals with the same
integral without restriction on the equator.
The functional integration is carried out on continuous functions.

\section{Calculation of the effective action}

We use the same trick as in \cite{ZJ} in order to get rid of the
functional integration. Namely, we introduce the "classical" 
solution $h_c$, and the new variable $\tilde{h}=h-h_c$.
Since the action is quadratic in the field we can write it in the following form:
\egy{S[ h ]=S[ h_c ]+S[ \tilde{h} ].}
The "classical" solution $h_c$ is the function that solves
the Laplace equation on the northern and southern hemisphere, respectively,
and on the equator satisfies the same boundary condition as the integration
variable in (\ref{funcint1}):
\egy{\triangle h_c(\vartheta,\varphi)=0 \mbox{ if } \vartheta\ne \frac{\pi}{2}, \quad
h_c\left(\frac{\pi}{2},\varphi\right)=h_0\left(\varphi\right).\label{Laplace}}
Changing the integration variable $h$ to $\tilde{h}$ we have:
\egy{P\left[ h_0 \right]=Z^{-1}e^{-S[h_c]}\int_
{\tilde{h}\left(\frac{\pi}{2},\varphi\right)=0}
\mathcal{D}\tilde{h}\,e^{-S[ \tilde{h} ]}\label{funcint2}}
Since the new boundary condition is $\tilde{h}\left(\frac{\pi}{2},\varphi\right)=0$,
the integral is now independent of $h_0$ so it can be absorped into $Z$.
\par
In the "classical" action we can perform an integration by parts:
\egy{S[ h_c ]=\frac{\sigma}{2}\int\limits_{0}^{2\pi}\mathrm{d}\varphi
\left\{\lim_{x\to -0}\left[ h_c \frac{\partial h_c}{\partial x}
\left(1-x^2\right)\right] - \lim_{x\to +0}
\left[ h_c \frac{\partial h_c}{\partial x}\left(1-x^2\right)\right]\right\}
\label{actionhc},}
where we changed to variables $x=\cos\vartheta$ and $\varphi$ for convenience.
\par
In order to find $h_c(x,\varphi)$ we have to solve two
Dirichlet-problems on the two hemispheres
with a common boundary condition at the equator. It is tempting to determine
the continuous and so squared integrable solution to (\ref{Laplace}) as an
expansion to spherical harmonics, which are eigenvectors of the Laplacian.
This way the effect of the Laplacian on the expansion could easily be
computed by changing the order of the sum and the Laplace operator.
However, the latter step is not always correct, the Laplacian not being
a bounded operator. We can convince ourselves that in the present case
changing the order of the infinite sum and the Laplacian yields erroneous
result. Instead, we use an expansion for which this step is allowed.
\par
In spherical polar coordinates the Laplace equation reads
\egy{\frac{\partial}{\partial x}(1-x^2)\frac{\partial h_c}{\partial x}
+\frac{1}{1-x^2}\frac{\partial^2 h_c}{\partial \varphi^2}=0.\label{Lapsph}}
Putting the usual Ansatz $h_c^n(x,\varphi)=Q^n(x)e^{in\varphi}$ into (\ref{Lapsph})
we obtain a second order equation for $Q^n$. For $n\ne 0$ we give the general
solution of this equation as the sum of the linearly independent solution of
the following two equations of first order:
\egy{\frac{dQ^n}{dx}=\frac{\pm n}{1-x^2}Q^n.}
The general solutions of (\ref{Lapsph}) are
\egy{
\begin{split}
h_c^n(x,\varphi) &= \left[a_n^+ \left(\frac{1-x}{1+x}\right)^
{\frac{\left| n \right|}{2}} + a_n^- \left(\frac{1+x}{1-x}\right)^
{\frac{\left| n \right|}{2}}\right]e^{in\varphi},\quad n \ne 0;\\
h_c^0(x,\varphi) &= a_0 + b \ln\frac{1-x}{1+x}.
\end{split}}
By prescribing regularity at the poles, only one term remains for each $n$:
\egy{
\begin{split}
h_c^n(x,\varphi) &= a_n z^{\left| n \right|}e^{in\varphi},\quad n \ne 0;\\
h_c^0(x,\varphi) &= a_0,
\end{split}}
where we introduced the notation
\egy{z=\sqrt{\frac{1-|x|}{1+|x|}}=\left\{
\begin{array}{l}
\tan\frac{\vartheta}{2},\quad \vartheta\le\frac{\pi}{2}\\
\cot\frac{\vartheta}{2},\quad \vartheta\ge\frac{\pi}{2}
\end{array}.\right.}
We are ready to give the solution of the two Dirichlet problems (\ref{Laplace}):
\egy{h_c(z,\varphi)=\frac{1}{\sqrt{2\pi}}\sum_{n=-\infty}^{\infty}\hat{h}_0(n)
e^{in\varphi} z^{\left|n\right|},\label{hc}}
where $\hat{h}_0(n)$ denote the Fourier coefficients of $h_0$.
If $\sum\limits_{n=-\infty}^\infty n^2 |\hat{h}_0(n)|^2<\infty$ then
\egy{\lim_{z\to 1}\frac{\partial h_c}{\partial z}(z,\varphi)=\frac{1}{2\pi}
\sum_{n=-\infty}^\infty |n| \hat{h}_0(n) e^{in\varphi}.}
Since
\egy{\frac{\partial h_c}{\partial x}=
-\frac{\mathrm{sgn}\,x}{(1+|x|)\sqrt{1-x^2}}\frac{\partial h_c}{\partial z},}
the effective action can be given in the following form:
\egy{S[h_c]=\sigma\int\limits_0^{2\pi}\mathrm{d}\varphi\,h_0(\varphi)
\lim_{z\to 1}z\frac{\partial h_c}{\partial z}(z,\varphi)=
2\sigma\sum_{n=1}^{\infty} n|\hat{h}_0(n)|^2.\label{1/f}}
where the last equality was obtained by using $\hat{h}_0(-n)=\hat{h}_0(n)^*$.
\par
Thus we have arrived at our main result: the action describing the probability
density of the height configurations on the equator is exactly the same
as the one characterizing Gaussian $1/f$ noise.

\subsection{Generalization to arbitrary circles}

In case the boundary values are prescribed not on the equator but for instance
on the circle at latitude angle $0<\vartheta_0<\pi$, we introduce similar
variables as before:
\egy{z=\sqrt{\frac{1-x}{1+x}} \quad z_0=\sqrt{\frac{1-x_0}{1+x_0}},}
where $x_0=\cos \vartheta_0$. The solutions of the two Dirichlet problems are
\egy{h_c^\pm (z,\varphi)=\frac{1}{\sqrt{2\pi}}\sum_{n=-\infty}^{\infty}\hat{h}_0(n)
e^{in\varphi} \left(\frac{z}{z_0}\right)^{\pm\left|n\right|},} 
where $h_c^+$ $(h_c^-)$ is defined for $\vartheta\leq\vartheta_0$
$(\vartheta\geq\vartheta_0)$.
\par
The integration in $S[ h_c ]$ is carried out along the circle at 
latitude angle $\vartheta_0$,
where the derivatives of the solutions are not continuous: 
\egy{S[ h_c ]=\frac{\sigma}{2}\int\limits_{0}^{2\pi}\mathrm{d}\varphi
\left\{\lim_{x\to -x_0}\left[ h_c^- \frac{\partial h_c^-}{\partial x}
\left(1-x^2\right)\right] - \lim_{x\to +x_0}
\left[ h_c^+ \frac{\partial h_c^+}{\partial x}\left(1-x^2\right)\right]\right\}
\label{actionhc2}.}
Since $(1-x^2)\frac{dz}{dx}=-z$, the $-z_0^{-1}$ factor introduced by 
differentiating $h_c^-$ with respect to $z$ is cancelled after carring out
the limit. The same is valid to the $h_c^+$ term with the opposite sign.
Therefore we arrive at exactly the same effective action (\ref{1/f})
as in the equatorial case.

\section{Final remarks}

The question naturally arises whether boundary conditions prescribed on
objects different from circles yield the effective action of the
Gaussian $1/f$ noise. A possible way of finding such objects is to apply
conformal transformations to the solution obtained in case of circles.
These transformations leave Laplace equation invariant but the
boundary curves where the solutions on the two regions of the sphere
are jointed might take more general shape. Because of the large
variety of conformal transformations, finding the ones that
result in the action of Gaussian $1/f$ noise requires more detailed analysis.
\par
It would be worth performing the calculation on other surfaces, for instance
on  a negative curvature surface besides the plane and the sphere. 
Furthermore, the generalization to surfaces of non-trivial topology
would be interesting as well.

\section*{Acknowledgements}
We thank Z. Horv\'ath and Z. R\'acz for helpful discussions.
This research has been partially supported by the Hungarian Academy
of Sciences (Grant No. OTKA T043734).

%\section*{References}

\end{document}